\documentclass[conference]{IEEEtran}
\IEEEoverridecommandlockouts
\usepackage{cite}
\usepackage{amsmath,amssymb,amsfonts}
\usepackage{algorithmic}
\usepackage{graphicx}
\usepackage{textcomp}
\usepackage{xcolor}
\usepackage{bm}
\usepackage{cuted}
\usepackage{algorithm}
\usepackage{algorithmic}
\usepackage{float}
\usepackage{balance}
\def\BibTeX{{\rm B\kern-.05em{\sc i\kern-.025em b}\kern-.08em
    T\kern-.1667em\lower.7ex\hbox{E}\kern-.125emX}}

\usepackage{fancyhdr}
\usepackage{titlesec}
\usepackage[T1]{fontenc} 
\usepackage[utf8]{inputenc}
\usepackage{scalefnt}

\definecolor{customcolor}{RGB}{209,76,43}

\newcommand{\calibrilight}{\fontfamily{Calibri Light}\selectfont}

\fancypagestyle{firstpage}{
    \fancyhf{}
    \fancyhead[L]{\textcolor{customcolor}{\scalefont{0.92}\calibrilight\textit{Paper Accepted in Late Breaking Research Publications in 2024 IEEE Energy Conversion Congress and Exposition (ECCE) }}}
    
}

\begin{document}

\title{From Black Box to Clarity: AI-Powered Smart Grid Optimization with Kolmogorov-Arnold Networks\\

}

\small 
\author{\IEEEauthorblockN{Xiaoting Wang}
\IEEEauthorblockA{\textit{Department of Electrical and} \\ \textit{Computer Engineering} \\
\textit{University of Alberta}\\
Edmonton, Canada \\
xiaotin5@ualberta.ca}
\and
\IEEEauthorblockN{Yuzhuo Li}
\IEEEauthorblockA{\textit{Department of Electrical and} \\ \textit{Computer Engineering} \\
\textit{University of Alberta}\\
Edmonton, Canada \\
yuzhuo@ualberta.ca}
\and
\IEEEauthorblockN{Yunwei Li}
\IEEEauthorblockA{\textit{Department of Electrical and} \\ \textit{Computer Engineering} \\
\textit{University of Alberta}\\
Edmonton, Canada \\
yunwei.li@ualberta.ca}
\and
\IEEEauthorblockN{Gregory Kish}
\IEEEauthorblockA{\textit{Department of Electrical and} \\ \textit{Computer Engineering} \\
\textit{University of Alberta}\\
Edmonton, Canada \\
gkish@ualberta.ca}
}
\normalsize

\maketitle
\thispagestyle{firstpage}

\begin{abstract}
This work is the first to adopt Kolmogorov-Arnold Networks (KAN), a recent breakthrough in artificial intelligence, for smart grid optimizations. To fully leverage KAN's interpretability, a general framework is proposed considering complex uncertainties. The stochastic optimal power flow problem in hybrid AC/DC systems is chosen as a particularly tough case study for demonstrating the effectiveness of this framework.
\end{abstract}


\section{Introduction}


Data-driven methods, particularly deep learning, are increasingly adopted to address smart grid optimization challenges, especially considering intermittency and variable nature within systems \cite{fioretto2020,pan2022deepopf,Huang2024}. Yet, their black-box nature limits the practical application in high-risk real-world scenarios. To tackle this fundamental issue, this paper presents an interpretable algorithm architecture designed to provide essential transparency when adopting AI into smart grid optimization.

The proposed framework leverages the Kolmogorov-Arnold Networks (KAN) architecture, a recent AI breakthrough \cite{liu2024kan}, to enhance decision-making in smart grids. This is the first application of the KAN method in energy systems, demonstrating its capability to provide physical/mathematical insights and better decisions in complex and dynamic environments. 

The core idea of KAN is fundamentally different from classical AI like multi-layer perceptions (MLP) \cite{hornik1989multilayer,cybenko1989approximation} (i.e., the building block of many deep learning methods), and embedded activation functions in the edges of computation graph \cite{liu2024kan,abueidda2024deepokan,li2024kolmogorov,bozorgasl2024wav}. With a layered structure, KAN features the universal approximation power just like ordinary deep learning, however, provides much greater transparency: with much less neurons, KAN can learn as accurately as (if not better than) MLP with 10-100 times more parameters, and each activation function has clear physical/mathematical significance to guide design and decision.


A critical area within smart grids is hybrid AC/DC systems, where accurate and rapid responses to fluctuations are essential but challenging due to unclear system models. Optimizing problems like optimal power flow (OPF) in hybrid AC/DC systems serves as a tough case study to demonstrate the effectiveness of our proposed frameworks.  \color{black}

\section{Interpretable AI-Powered Smart Grid Optimization}
\subsection{Mathematical Formulation of Smart Grids Optimization}

 Due to the existence of uncertainties in smart grids, all optimization variables, and objective values are regarded as random variables, which requires adjusting system operating points in response to the uncertain changes \cite{Tang2017,Xiang2021}. 
 Let $\bm{\xi}$ denote the intermittency from renewable energy or loads (uncertainty sources like solar irradiation or load variation). Then, a general optimization formulation for smart grid tasks can be represented as 
\cite{Ergun2019,Xiang2021,roald2023power}: 
%
\begin{equation}
\label{eq:POPF}
    \min F(\bm{u},\bm{\xi}), \quad 
    \mathrm{s.t.}  \bm{f}(\bm{x},\bm{u},\bm{\xi}) = 0, \bm{h} (\bm{x},\bm{u},\bm{\xi}) \leq 0 
\end{equation}
%
%
where $F(\bm{u},\bm{\xi})$ denotes the objective function (e.g., generation costs, power losses, planning costs), $\bm{u}$ denotes the optimization variables (e.g., generation real outputs $P_g$),  
$\bm{x}$ denotes the state variables. Particularly, constraints in \eqref{eq:POPF} include the physical and network limits for the smart grid (e.g., voltage limits, thermal limits, power balance, and AC/DC conversion constraints). 

\subsection{Proposed KAN-based Framework}
%
%
%
\begin{figure*}[]
\centering
\includegraphics[width=1\textwidth]{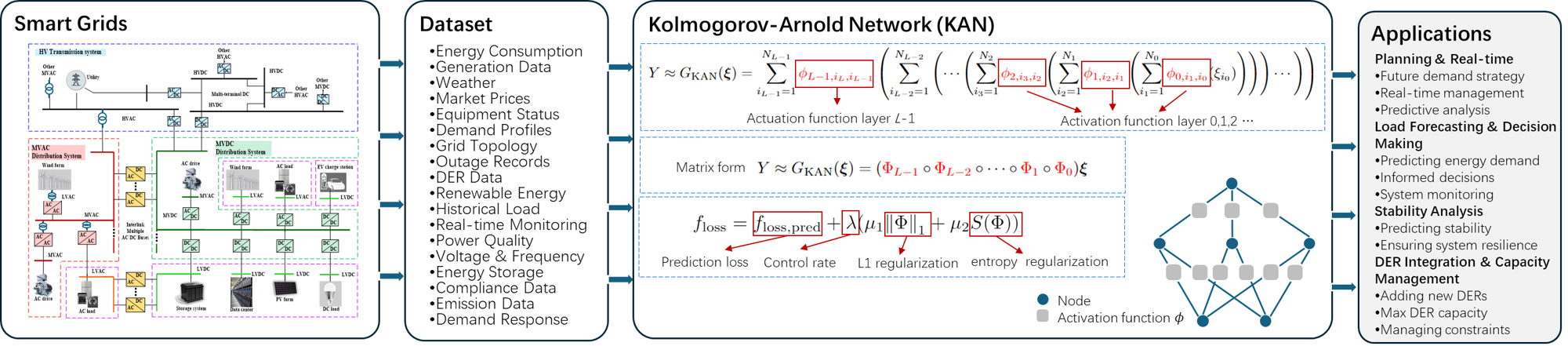}
\caption{The KAN-based framework for smart grid optimization tasks \cite{li2022smart,liu2024kan}}
\label{fig:kan_frame}
\end{figure*}

The KAN-based framework is shown in detail in Fig. \ref{fig:kan_frame}. Various smart grids operation data can be selected and fed into the KAN training process. Specially, the KAN is in the central role and its structure is generalized to adjust all the internal parameters to capture and learn the target distributions of the response of interests (e.g., generator outputs, bus voltages, power flow)
and serves as the digital replica of the smart grid optimization model for downstream critical applications (e.g., planning, forecasting, monitoring, etc.).


Let $Y = G(\bm{\xi})$ denote an optimization model 
 (e.g., economic dispatch \cite{Wang2022ED,Hu2021}\color{black}, OPF \cite{wang2024efficient,Liu2020}), where $Y$ represents the output of the model and could be the objective values and solutions of the optimization model. $\bm{\xi} =[\xi_1,\cdots,\xi_{N_0}] \in \mathbb{R}^{N_0}$ is a multivariate input vector with bounded domain (e.g., uncertain source from renewable energy sources (RES) and loads).
 According to the Kolmogorov-Arnold representation theorem \cite{kolmogorov1961representation}, 
 the optimization (e.g., stochastic OPF) model can be represented by a generalized KAN network with $L$ layers \cite{liu2024kan} (see Fig. \ref{fig:kan_frame}). 
\color{black} Specially, the learnable activation function offers good interpretability in terms of complex physical/mathematical models with different complexity levels. For example, activation functions can be a weighted summation of a basis function like spline function \cite{liu2024kan}, or Gaussian radial basis functions \cite{abueidda2024deepokan}. The loss function can be designed by considering task-specific performance. 
Readers can refer to  \cite{liu2024kan} for more implementation details.  

\section{Case Studies -- Stochastic OPF Application}
\begin{figure}[ht]
\centering
\includegraphics[width=0.5\textwidth]{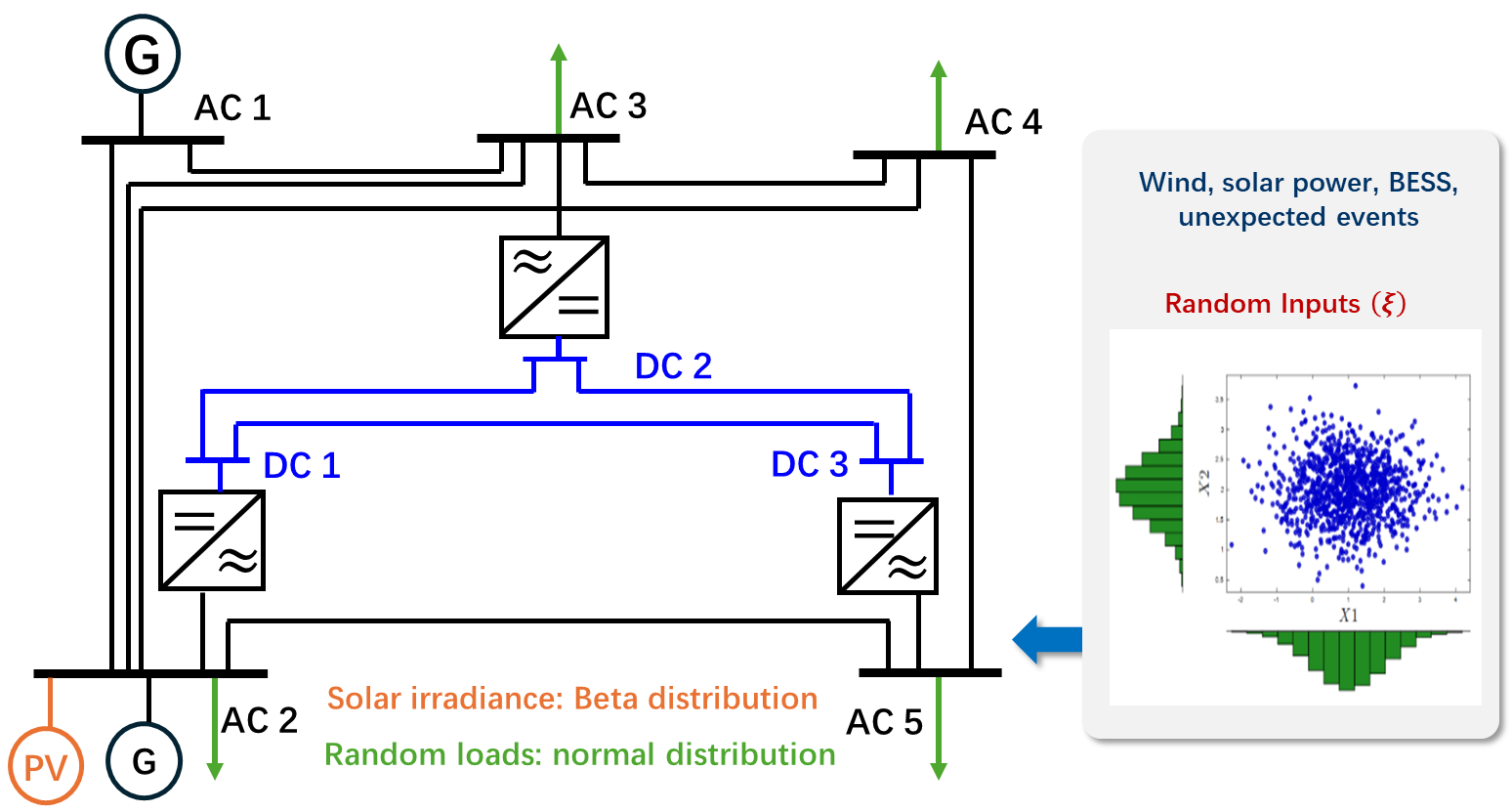}
\caption{Hybrid AC/DC system case study. single line diagram of the modified 5-bus test system. The blue lines indicate DC lines \cite{Ergun2019}.}
\label{fig:case5_diagram}
\end{figure}

In this section, we validate the effectiveness of the KAN framework for hybrid AC/DC systems in solving the stochastic OPF (SOPF) problem. Particularly, we conduct simulations using the modified IEEE 5-bus test system. The configuration of this system is illustrated in Fig. \ref{fig:case5_diagram} \cite{Beerten2012,Ergun2019}. 
All simulations have been conducted on a PC equipped with Intel Core i7-8700 (3.20GHz), 16GB RAM. \color{black} The SOPF for hybrid AC/DC is solved based on Julia and the KAN framework is constructed using pykan package \cite{liu2024kan}.

\begin{figure}[ht]
\centering
\includegraphics[width=0.5\textwidth]{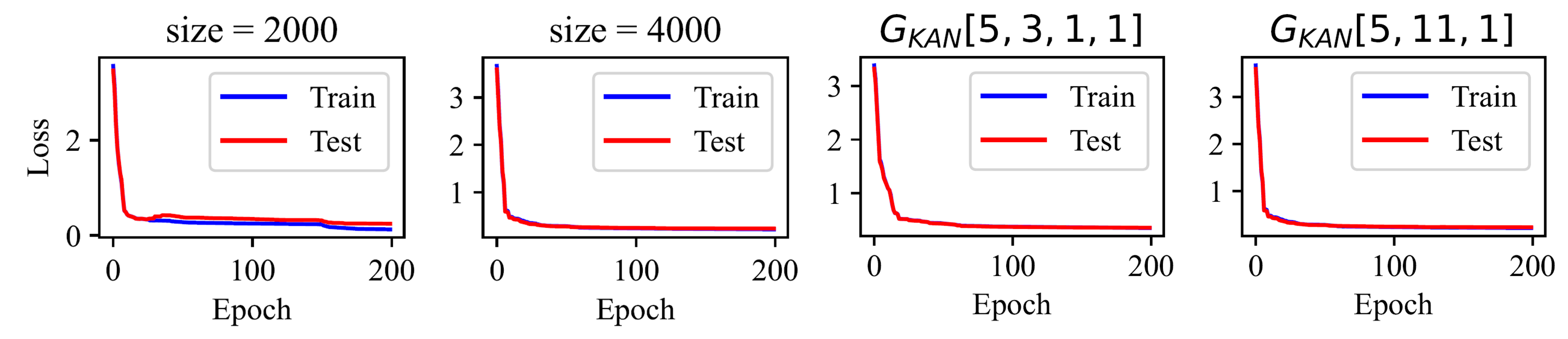}
\caption{Training and test loss for different sample sizes and KAN configurations. For different KAN configurations, the training sample size is set as 4000. 
}
\label{fig:training_test_loss}
\end{figure}

\begin{figure}[htbp]
\centering
\vspace{-0.3in}
\includegraphics[width=0.5\textwidth]{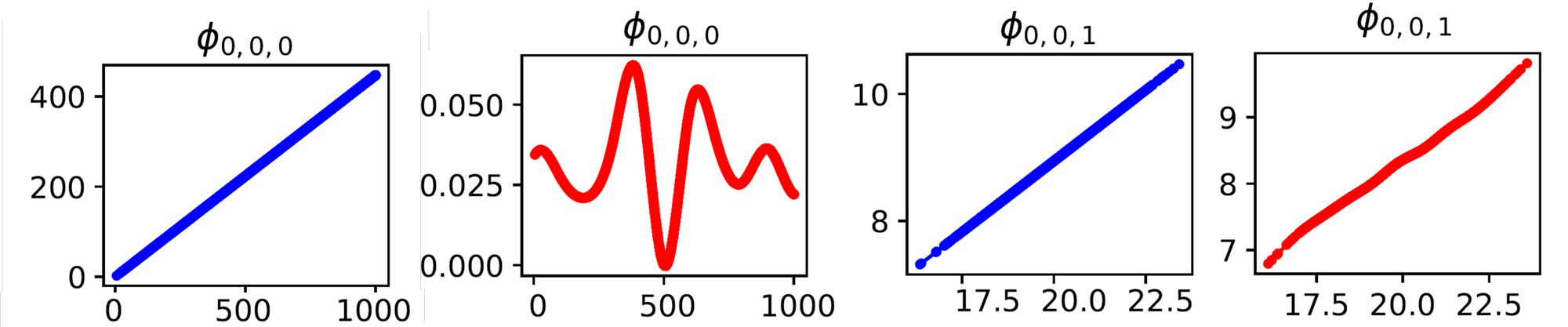}
\caption{Examples of activation function in for layer $l=0$ before and after training. Blue: before. Red: after.}
\label{fig:activation_function}
\end{figure}

\begin{figure}[htbp]
\centering
\includegraphics[width=0.5\textwidth]{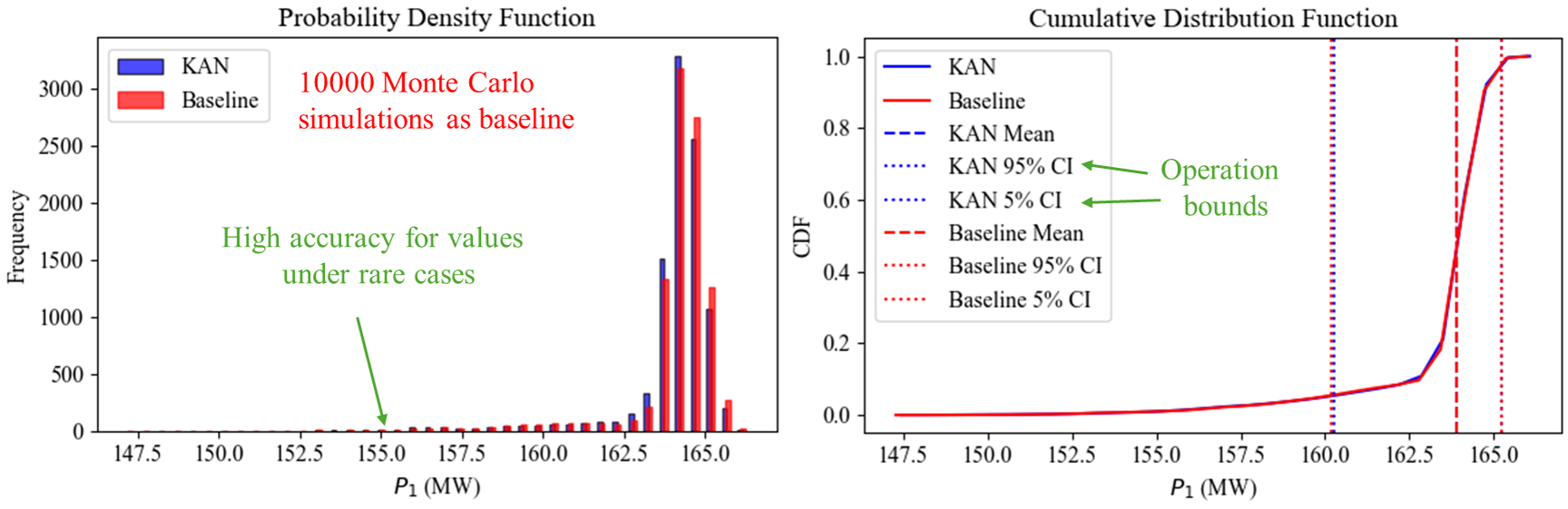}
\caption{The probability distribution functions (PDFs) and cumulative distribution functions (CDFs) of estimated SOPF solutions: generator output $P_{g_1}$. CI: confidence interval. 
}
\label{fig:histogram}
\end{figure}

Fig. \ref{fig:training_test_loss} shows the comparison of training and test losses for various training sample sizes and KAN network configurations, differentiated by the number of neurons. The results indicate that a three-layer KAN network, enabling non-smooth inner functions, can provide accurate estimations. 
Fig. \ref{fig:activation_function} presents activation function examples at layer $l=0$ before and after training. Functions inside KAN are observed with learnt non-linearity and capture the 
statistical features and correlations of uncertain inputs (e.g., solar power and load dynamics/distributions) in data\color{black}. This can be further validated through the statistical information on optimal solutions determined by KAN framework. Fig. \ref{fig:histogram} compares distribution functions obtained from the KAN and baseline (Monte Carlo simulations \color{black} \cite{zhang2010probabilistic}\color{black}). This includes the probability density functions (PDFs) and cumulative distribution functions (CDFs) of decision variables such as generator output $P_{g_1}$ for the SOPF problem. These results demonstrate that the KAN can capture values even under rare cases (e.g., line outages \cite{Wang2021}), facilitating reliable decision-making in hybrid AC/DC systems.
For example, system operators can identify typical operating ranges and extreme values according to the generator outputs. The calculated mean and variance help in guiding operating points adjustment. The calculated confidence intervals further provide probabilistic bounds of expected outputs. All these factors ensure systems are within safe operation ranges.

%

%

\section{Conclusions}

Being the first to adopt the AI-domain late breakthrough KAN in energy systems, this paper introduced the KAN-based framework for smart grid optimization under complex uncertainties. The proposed framework fully utilized the silent features of KAN in supporting general grid optimization tasks. Numerical studies on a practical optimization task, the SOPF problem in a hybrid AC/DC system, validated its effectiveness.  

\newpage
\balance
\bibliographystyle{IEEEtran}
\bibliography{IEEEabrv,HybridACDC.bib}

\end{document}